\documentclass[%
 reprint,
preprintnumbers,
 amsmath,amssymb,
 aps,
 prl,
floatfix
]{revtex4-1}

\usepackage{color}

\usepackage[colorlinks=true,linkcolor=blue,citecolor=blue, urlcolor=blue]{hyperref}   
\usepackage{graphicx}
\usepackage{dcolumn}
\usepackage{bm}

\newcommand{\kompost}{K{\o}MP{\o}ST}

\newcommand{\tauhydro}{\tau_\text{hydro}}
\def\Eq#1{Eq.~(\ref{#1})}

\usepackage{footmisc}

\def\Eqs#1{Eqs.~(\ref{#1})}

\def\eqs#1{(\ref{#1})}

\def\Fig#1{Fig.~\ref{#1}}

\def\prlsection#1{{\it #1}.---}

\def\aprlsection#1{\section{Supplemental material\label{sec:supplemental}}}

\def\App#1{the \hyperref[sec:supplemental]{supplemental material}}

\def\x{{\mathbf x}}

\def\wt{\tilde{w}}
\def\C{C_\infty}

\newcommand{\etas}{\eta_s}

\begin{document}

\preprint{INT-PUB-19-036}

\title{Hydrodynamic attractors, initial state energy and particle production \\ in relativistic nuclear collisions}
\author{Giuliano Giacalone}
\email[]{giuliano.giacalone@ipht.fr}
\affiliation{Institut de physique th\'eorique, Universit\'e Paris Saclay, CNRS, CEA, F-91191 Gif-sur-Yvette, France}

\author{Aleksas Mazeliauskas}
\email[]{aleksas.mazeliauskas@cern.ch}
\affiliation{Theoretical Physics Department, CERN, CH-1211 Gen\`eve 23, Switzerland}
\affiliation{Institut f\"ur Theoretische Physik, Universit\"at Heidelberg, Philosophenweg 16, D-69120 Heidelberg, Germany}

\author{S\"{o}ren Schlichting}
\email[]{sschlichting@physik.uni-bielefeld.de}
\affiliation{Fakult\"{a}t f\"{u}r Physik, Universit\"{a}t Bielefeld, D-33615 Bielefeld, Germany}

\date{\today}

\begin{abstract}
We exploit the concept of hydrodynamic attractors to establish a macroscopic description of the early-time out-of-equilibrium dynamics of high energy heavy-ion collisions. One direct consequence is a general relation between the initial state energy and the produced particle multiplicities measured in experiments. 
When combined with an ab initio model of energy deposition, the entropy production during the pre-equilibrium phase naturally explains the universal centrality dependence of the measured charged particle yields in nucleus-nucleus collisions. We further estimate the energy density of the far-from-equilibrium initial state and discuss how our results can be used to constrain non-equilibrium properties of the quark-gluon plasma.

\end{abstract}

\maketitle

\prlsection{Introduction}
Understanding the equilibration of isolated quantum systems is a fundamental
question that touches physical phenomena across vastly different energy scales,
from micro kelvin temperatures in cold atom experiments to trillion kelvin
temperatures in the dense strong-interaction matter produced in ultra-relativistic nuclear
collisions~\cite{Busza:2018rrf, Schafer:2009dj, Adams:2012th, Schlichting:2019abc}. One outstanding discovery made in the field of heavy-ion collisions is that the system created about $1\, \rm{fm}/c$ ($\approx 3\cdot 10^{-24}\,\text{s})$ after the impact of two relativistic nuclei can be described as a deconfined plasma of quarks and gluons (QGP) with
macroscopic properties of temperature and velocity~\cite{Adams:2005dq,Adcox:2004mh,Back:2004je,Arsene:2004fa}. Such ``unreasonable
effectiveness of hydrodynamics'' in describing the violent expansion of the QGP droplets
triggered a new research area in mathematical physics devoted to the study of \emph{hydrodynamic
attractors}, that emerge in out-of-equilibrium systems experiencing very fast memory loss
of initial conditions and exhibiting a universal approach towards thermal equilibrium~\cite{Florkowski:2017olj, Romatschke:2017ejr}.

In this article we show
that hydrodynamic attractors can be used to describe entropy production in relativistic nuclear collisions and to make robust
estimates of initial-state energy \emph{before} the onset
of equilibration.
We derive a simple formula, \Eq{eq:Entropy}, that relates the energy density of the initial state to the measured charged
particle multiplicity, $dN_{\rm ch}/d\eta$, 
and point out two important phenomenological consequences of this result.
We show that the universal centrality dependence of $dN_{\rm ch}/d\eta$ across a wide
range of collision systems can be naturally reproduced by combining
the initial-state energy deposition in high-energy quantum chromodynamics (QCD) with the non-linear entropy production during the equilibration process.
Secondly, we determine the initial energy per unit space-time rapidity,
$dE_0/d\etas$, for different collision centralities at the Relativstic Heavy-Ion Collider (RHIC) and the Large Hadron Collider (LHC). 
By comparing our results to the experimentally measured $dE_\text{final}/dy$ in the final state, we estimate the work performed during the expansion of the system~\cite{Gyulassy:1983ub}
and discuss how such an analysis constrains non-equilibrium and transport properties of the QGP.

\prlsection{Hydrodynamic attractors \& Entropy production} We describe 
the early time dynamics ($\tau\lesssim 1\,\text{fm}/c$) of
the high-temperature QCD plasma created in high-energy heavy-ion collisions by
the out-of-equilibrium evolution of a boost-invariant and transversely homogeneous conformal system 
~\cite{Bjorken:1982qr}.
Energy-momentum conservation dictates that the energy density, $e=T^{\tau\tau}$, evolves according to
\begin{equation}
\label{eq:EnergyConservation}
    \partial_\tau e = - \frac{e + P_L}{\tau}\;,
\end{equation}
where  $P_L\equiv\tau^2T^{\etas\etas}$ is
the longitudinal pressure and we use proper time $\tau \equiv\sqrt{t^2-z^2}$ and space-time rapidity $\etas\equiv \text{atanh}\, z /t$ coordinates.
In (local) thermal equilibrium, the longitudinal
pressure is directly related to the energy density via an equation of state,
e.g. $P_L=e/3$ for a conformal system. While for small deviations around equilibrium the longitudinal pressure
is determined by hydrodynamic constitutive relations in terms of the gradient
expansion 
$P_L/e=\frac{1}{3}-\frac{16}{9}\frac{\eta/s}{\tau T}$, where $\eta /s$ is the specific shear viscosity~\cite{Florkowski:2017olj, Romatschke:2017ejr},
this is generally not
the case far from equilibrium, where, for instance, at early times after the
collision of heavy nuclei the system is highly anisotropic $P_L \ll e$~\cite{Schlichting:2019abc}. Nevertheless, new insights from
microscopic equilibration studies%
~\cite{Heller:2011ju,Heller:2015dha,Romatschke:2017vte, Heller:2016rtz,Strickland:2017kux,Blaizot:2017ucy,Strickland:2018ayk,Behtash:2019txb, 
Behtash:2017wqg,Behtash:2018moe,
Kurkela:2015qoa,Keegan:2015avk, Keegan:2016cpi, Kurkela:2018vqr,Kurkela:2018wud,Kurkela:2018xxd,Kurkela:2018oqw, Kurkela:2019kip} 
point to the 
existence of a hydrodynamic attractor~\cite{Heller:2015dha}, where the far-from-equilibrium system
displays an effective constitutive equation $P_L/e=f(\wt)$ well before
reaching local thermal equilibrium.
Such attractor behavior has been established for a
number of different microscopic theories (QCD Kinetic Theory~\cite{Kurkela:2018vqr,Kurkela:2018wud,Kurkela:2018xxd,Kurkela:2018oqw}, Boltzmann RTA~\cite{Heller:2016rtz,Strickland:2017kux,Blaizot:2017ucy,Strickland:2018ayk,Behtash:2019txb} and
AdS/CFT~\cite{Heller:2011ju,Heller:2015dha,Romatschke:2017vte}), where the time evolution on the attractor
is controlled by a single scaling variable, $\wt={\tau T_{\rm eff}(\tau)
}/{(4 \pi\eta/s)}$, where $T_{\rm eff}(\tau)$ is  an effective
temperature
such that $e(\tau)\equiv\frac{\pi^2}{30} \nu_{\rm eff} T_{\rm eff}^4(\tau)\;$
($\nu_\text{eff}$ is the number of effective  degrees of freedom, e.g.,\ $\nu_\text{eff}=16$ for ideal gluonic gas).

\begin{figure}
    \centering
    \includegraphics[width=\columnwidth]{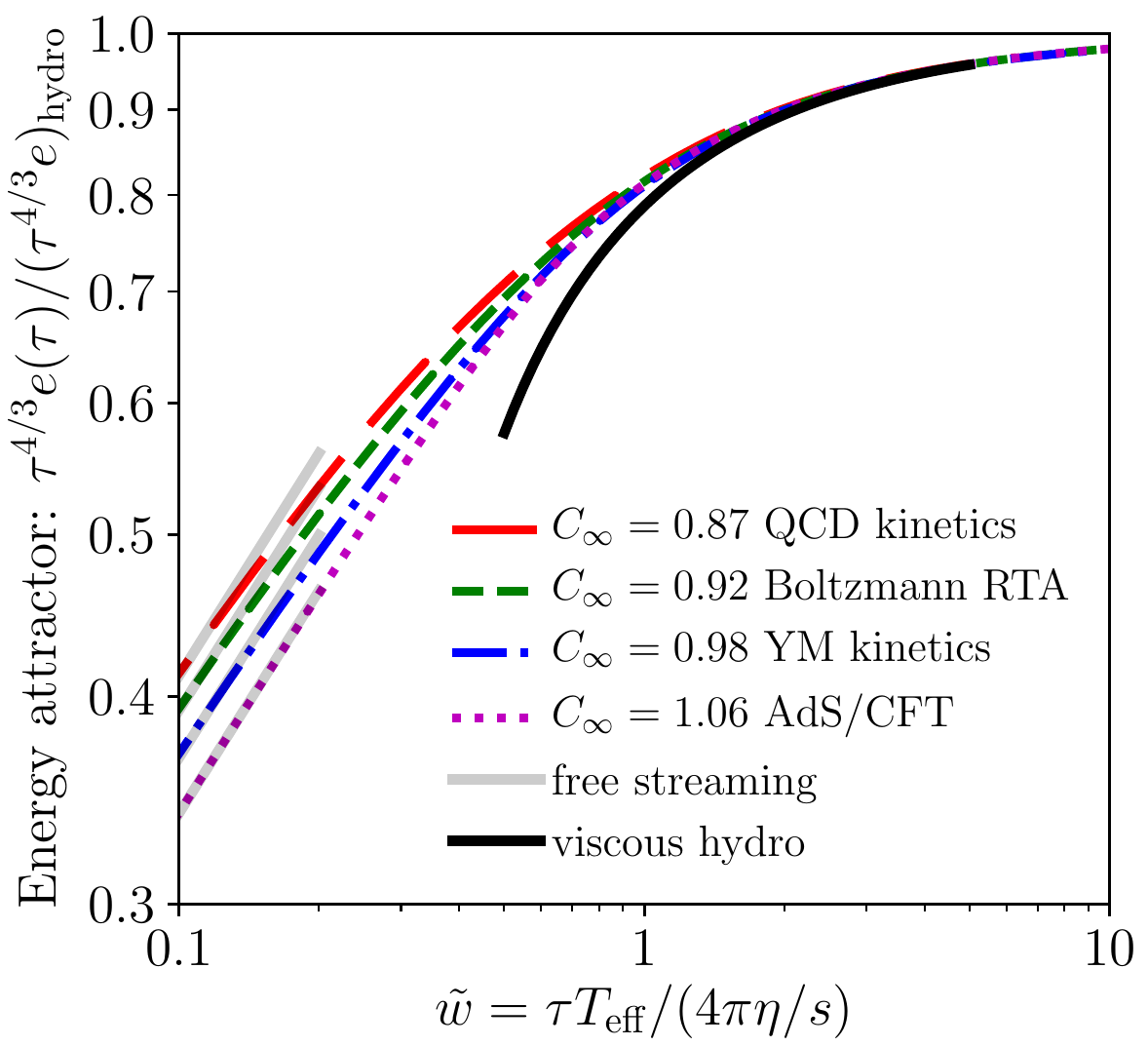}
    \caption{Hydrodynamic attractor for pre-equilibrium evolution of the energy density obtained from QCD and Yang Mills (YM) kinetic theory~\cite{Kurkela:2018vqr,Kurkela:2018wud,Kurkela:2018xxd,Kurkela:2018oqw}, AdS/CFT~\cite{Heller:2011ju,Heller:2015dha,Romatschke:2017vte} and Boltzmann
RTA~\cite{Heller:2016rtz,Strickland:2017kux,Blaizot:2017ucy,Strickland:2018ayk,Behtash:2019txb}.
Solid lines show the asymptotic behavior of the attractor curves given by 
\Eq{eq:AttractorLimits}.
 \label{fig:attr}}
\end{figure}

Based on these insights, the conservation law in \Eq{eq:EnergyConservation} can
be integrated, yielding a universal relation between the initial state energy
density $e_0$ at very early times $\wt(\tau_0)\ll 1$, and the energy density
$e(\tau_\text{hydro})$ of the near thermal system at later times
$\wt(\tau_\text{hydro})\gg 1$
\begin{eqnarray}
\label{eq:EnergyEvolution}
e(\tau_\text{hydro}) = e_0 \exp\left( -\int_{\wt_0}^{\wt_\text{hydro}} \frac{d\wt}{\wt} \frac{1+f(\wt)}{\frac{3}{4}-\frac{1}{4}f(\wt)} \right)\;.
\end{eqnarray}
Close to equilibrium $f(\wt_\text{hydro})\approx 1 /3$ and the energy density of the
longitudinally expanding plasma follows the Bjorken scaling $e(\tau)= e_{\rm
hydro} \left(\tau / \tauhydro\right)^{-4/3}$,
while the entropy density per
unit rapidity, $s\tau$, remains constant~\cite{Bjorken:1982qr}. 
Eventually, for  $\tau\gtrsim R/c$, where $2R$ denotes the transverse extent of the system,
the QGP fireball starts expanding in the transverse plane and  ultimately freezes out in
color neutral hadrons~\cite{Teaney:2009qa}.
During the transverse expansion the QGP remains close to equilibrium and 
the total entropy per unit rapidity ${dS}/{d\etas}=A_{\bot}~(s\tau )_{\rm hydro}$ (where $A_{\bot}=\pi R^2$)
 is approximately conserved onwards from the time $\tauhydro$ when the QGP can be described as an almost ideal fluid.
Ultimately, on the freeze-out surface ${dS}/{d\etas}$ becomes
proportional to the produced charged hadron multiplicity,
$dN_\text{ch}/d\eta$. The multiplicity of final-state
particles emitted from the QGP is therefore a sensitive probe of the entropy
production during the pre-equilibrium phase.

Strikingly, the correspondence between initial-state energy density and charged
hadron multiplicity can be quantified further using the theory of hydrodynamic
attractors. 
By factoring out the late time Bjorken scaling
from 
\Eq{eq:EnergyEvolution} the evolution of the
energy density  during the pre-equilibrium phase can be
characterized by an attractor curve $\mathcal{E}(\wt)$
\begin{eqnarray}
\label{eq:EnergyAttractor}
\frac{e(\tau)\tau^{4/3}}{e_{\rm hydro}\tauhydro^{4/3}} = \mathcal{E}\left(\wt=\frac{T_{\rm eff}(\tau) \tau}{4 \pi\eta/s}\right)\;.
\end{eqnarray}
As can be seen from
\Fig{fig:attr}, the function $\mathcal{E}(\wt)$ smoothly interpolates between an early
free-streaming and late-stage viscous hydrodynamics~\cite{Romatschke:2017vte,Kurkela:2018vqr}
\begin{align}
\label{eq:AttractorLimits}
\begin{split}
  \mathcal{E}(\wt \ll 1) &= \C^{-1} \wt^{4/9} \qquad (\text{free streaming})\;, \\
  \mathcal{E}(\wt \gg 1) &= 1 -\frac{2}{3 \pi\wt} \qquad\;\;(\text{viscous hydro})\;,
\end{split}
\end{align}
where $\C$ is a constant of order unity.  
Even though the evolution at intermediate times can be different for different
microscopic theories, the overall similarity between different theories is
remarkable. Most importantly for our purpose, all curves have the same universal
characteristics, \Eq{eq:AttractorLimits},  at early and late times, irrespective of the underlying
microscopic theory.

Based on \Eq{eq:EnergyAttractor}, we can immediately establish a quantitative
relation between the energy densities $e(\tau)$ at various stages, which upon use of the
thermodynamic relations $Ts=e+p$ and $p=e /3$ once the system is close to
equilibrium turns into an estimate of the entropy density per unit rapidity
\begin{eqnarray}
  (s\tau)_{\rm hydro} = \frac{4}{3} \left( \frac{\pi^2}{30} \nu_{\rm eff}\right)^{1/4} \Bigg(\lim_{\tau \to 0} \frac{e(\tau)\tau^{4/3}}{\mathcal{E}\big(\frac{T_{\rm eff}(\tau) \tau}{4 \pi\eta/s}\big)}\Bigg)^{3/4}\!\!.
\end{eqnarray}
Evaluating the limit according to \Eq{eq:AttractorLimits} one arrives at the central result of this paper, namely the relation
\begin{eqnarray}
\label{eq:Entropy}
(s\tau)_{\rm hydro} &&= \frac{4}{3} \C^{3 /4} \left(4\pi \frac{\eta}{s}\right)^{1/3} \left( \frac{\pi^2}{30} \nu_{\rm eff}\right)^{1/3} \left(e\tau\right)^{2/3}_{0},
\end{eqnarray}
from which one can directly estimate the charged particle multiplicity as discussed above:
\begin{eqnarray}
\label{eq:Multiplicity}
\frac{dN_{\rm ch}}{d\eta} \approx \frac{1}{J} A_{\bot}~(s\tau)_{\rm hydro}\frac{N_{\rm ch}}{S}\;. 
\end{eqnarray}
Here $S/N_\text{ch}\equiv \left(dS /dy \right) / \left( dN_\text{ch} /d y \right)  \approx 6.7\text{--}8.5$ is the entropy per charged particle at
freeze-out~\cite{Hanus:2019fnc} and $J\approx 1.1$ is a Jacobian factor
between particle rapidity $y$ and pseudo rapidity $\eta$~\cite{Adam:2016thv}.

Equations
\eqs{eq:Entropy} and
\eqs{eq:Multiplicity} establish a one-to-one correspondence between the
initial-state energy per unit rapidity $dE_{0}/d \etas \approx A_{\bot}
(e\tau)_0$ and the charged particle multiplicity ${dN_{\rm ch}}/{d\eta}$. One crucial feature 
of this result is that it accounts for the entropy production during the pre-equilibrium phase,
which gives rise to a nontrivial dependence on the initial-state energy
density $(e \tau)^{2 /3}_0$ as well as on the transport coefficient $(\eta/s)^{1 /3}$.
Our estimate includes all relevant pre-factors, in particular, the  
constant $\C$, which is the property of the hydrodynamic attractor, \Eq{eq:EnergyAttractor},
and depends on the microscopic physics of equilibration. However it is
striking to observe that for the different theories shown in
\Fig{fig:attr} the variation of $\C$ is only at $\sim 10\%$ level.
We emphasize that \Eq{eq:Entropy} is entirely based on the macroscopic evolution described by a generic hydrodynamic attractor, which sets it apart from previous parametric estimates of entropy production based on particular microscopic scenarios~\cite{Baier:2002bt,Baier:2011ap,vanderSchee:2015rta,Berges:2017eom}.

\begin{figure}
    \centering
    \includegraphics[width=\linewidth]{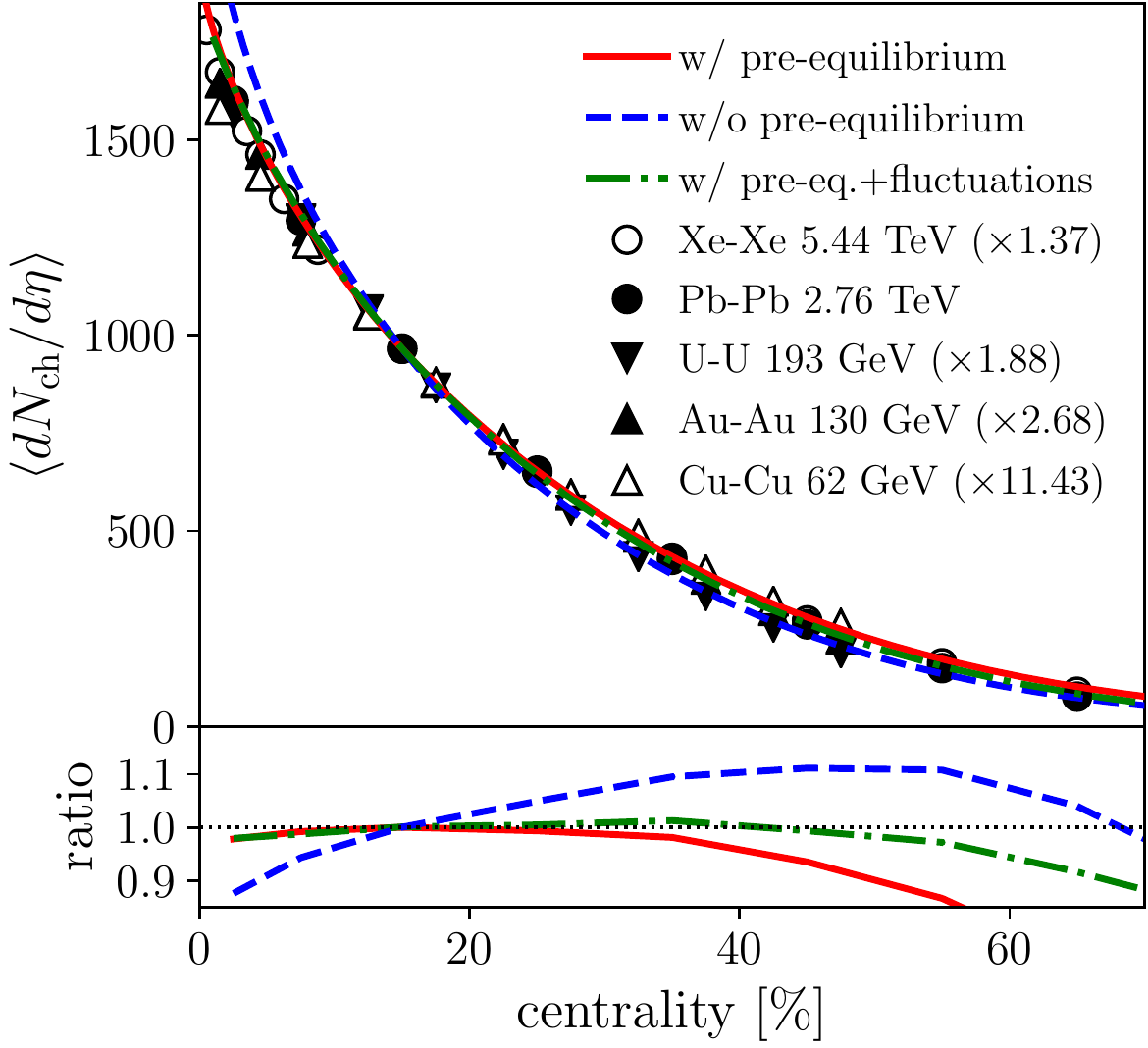}  
    \caption{ 
    The effect of pre-equilibrium dynamics on the centrality dependence of the charged-particle multiplicity, $dN_{\rm ch}/d\eta$,  can be appreciated by comparing the theoretical estimate on the gluon number (blue dashed line), given by Eq.~(\ref{eq:woPre}), to that on the charged-particle multiplicity after equilibration (red solid line), given by Eq.~(\ref{eq:wPre}). The green dot-dashed line includes as well fluctuations in the initial state energy within a Glauber Monte Carlo approach. Experimental data points are shown for different collision systems: Xe-Xe~\cite{Acharya:2018hhy}, Pb-Pb~\cite{Aamodt:2010cz}, U-U~\cite{Adare:2015bua}, Au-Au~\cite{Alver:2010ck}, and Cu-Cu~\cite{Alver:2010ck} collisions. All
curves are normalized to present the same value of multiplicity as ALICE Pb-Pb
data in the 10-20\% centrality bin. The bottom panel shows the ratio between 2.76 TeV Pb-Pb data and the theory curves.
}
    \label{fig:Nvscent}
\end{figure}
\prlsection{Centrality dependence of particle multiplicity}One important phenomenological consequence of the entropy production in the
pre-equilibrium phase concerns 
the determination of
initial conditions for
hydrodynamic simulations of heavy ion collisions (see e.g.\ \cite{Heinz:2013th}). 
While strictly
speaking our estimate of the entropy density in \Eq{eq:Entropy} was derived
assuming a one dimensional expansion,
the influence of transverse gradients can be
neglected over the short pre-equilibrium times and we can directly promote \Eq{eq:Entropy} 
to an estimate for the \emph{local} entropy density,
$\left.\tau s(\tau,\mathbf {x}_{\bot})\right|_{\tau=\tauhydro}$. %
  Specifically, the pre-flow $v_{\bot}\sim \tauhydro\mathbf{\nabla}_{\bot}e/e$ is negligible as long as gradients $\mathbf{\nabla}_{\bot}e/e$ are small on the scale of hydrodynamization time $(c \tauhydro)$ \cite{Keegan:2016cpi, Kurkela:2018vqr} and the one dimensional constitutive relation for $P_L/e$ approximately remains valid, as explicitly confirmed in~\cite{Kurkela:2020sal}.
Effectively, \Eq{eq:Entropy}  then provides a non-linear map of the initial-state energy density profile to the entropy density profile at later times $\tau \sim \tauhydro$.
 
  Now, in order to illustrate the impact of the pre-equilibrium phase, we will
 study the effects on the centrality dependence of the charged particle
 multiplicity, within a simple initial state model based on the color-glass
 condensate effective theory of high-energy QCD \cite{Gelis:2010nm}.
 Within the dilute-dense formulation of this theory~\cite{Dumitru:2001ux,Lappi:2006hq,Blaizot:2004wu,Blaizot:2008yb,Blaizot:2010kh}, both the initial energy
 density per unit rapidity $(e\tau)_0$ as well as the initial gluon multiplicity per unit
 rapidity $(n\tau)_0$ can be calculated from $\mathbf{k}_{\bot}$-factorization, and are
 given in terms of convolutions of unintegrated gluon distributions
 \footnote{We note that explicit comparisons of dilute-dense and
  dense-dense calculations in Ref.~\cite{Blaizot:2010kh} show that in the momentum range $p_T
  \gtrsim Q_s$, which gives the dominant contribution to ${dE}/{dy}$, the
  dilute-dense $k_T$ factorization calculations accurately describe dense-dense
Classical Yang-Mills results.}.
 Essentially, one finds that (up to logarithmic corrections) 
 $(e\tau)_0$ and $(n\tau)_0$
 are proportional to the (local) saturation scales
 $Q_{s}(\mathbf{x}_\bot)$ of the two colliding nuclei%
~\cite{Dumitru:2001ux,Lappi:2006hq}:
\begin{eqnarray}
  \label{eq:e0}
  (e\tau)_0(\mathbf{x}_\bot) &\propto& (Q_{s}^{<})^2(\mathbf{x}_\bot) Q_{s}^{>}(\mathbf{x}_\bot)\;, \\
  (n\tau)_0(\mathbf{x}_\bot) &\propto& (Q_{s}^{<})^2(\mathbf{x}_\bot), \label{eq:n0}
\end{eqnarray}
where $Q_{s}^{> / <}$ is the saturation scale of the nucleus representing larger/smaller $Q_s$ at position $\mathbf{x}_\bot$.

Since the saturation scale locally characterizes the longitudinally integrated density of color charge inside the nucleus, it is generically proportional to the nuclear thickness
\begin{eqnarray}
  \label{eq:QsTA}
  Q_s^2(\mathbf{x}_\bot) \propto T(\mathbf{x}_\bot),
\end{eqnarray}
whose definition is recalled in \App{app}~\footnote{\label{suppl}The supplemental material provides the description of the Monte Carlo Glauber model of nucleus-nucleus collisions used to determine the nuclear thickness functions, which are  needed for the calculation of the initial-state energy profile. In addition, we provide detailed formulas for calculating the centrality dependence of the averaged charged-particle multiplicity and the initial-state energy per rapidity, i.e. the lines and shaded bands in Figs. 2 and 3 of the manuscript. The supplemental material includes additional Refs.~\cite{DeJager:1987qc,Albacete:2014fwa}}. Based on these considerations,  one can then try to estimate the charged particle multiplicity per unit rapidity from the initial gluon multiplicity $(n\tau )_0$ (i.e.\ w/o pre-equilibrium) according to
\begin{eqnarray}
\label{eq:woPre}
\frac{dN_{\rm ch}}{d \eta} \propto \int d^2\mathbf{x}_\bot~ T^{<}(\mathbf{x}_\bot)\;,
\end{eqnarray}
as was done, for example, in \cite{Mace:2018vwq,Schenke:2012fw}. However, such an
estimate is appropriate only when there is no significant amount of particle
production in the final state. Conversely, if the initial state evolves into an
almost ideal QGP fluid, one needs to account for the entropy production during
the pre-equilibrium phase. By employing \Eqs{eq:Entropy} and \eqs{eq:Multiplicity} 
the  charged particle multiplicity is then estimated from the
initial state energy density $(e \tau)_0$, (i.e.\ w/ pre-equilibrium) as
\begin{eqnarray}
\label{eq:wPre}
\frac{dN_{\rm ch}}{d \eta} \propto \int d^2\mathbf{x}_\bot~ \left( T^{<}(\mathbf{x}_\bot) \sqrt{T^{>}(\mathbf{x}_\bot)} \right)^{2/3} \;.
\end{eqnarray}
We illustrate the difference between the two estimates in the upper panel of \Fig{fig:Nvscent}, where we
compare the centrality dependence of the multiplicity ${dN_{\rm ch}}/{d \eta}$
from Eq.~(\ref{eq:wPre}) (solid line) and Eq.~(\ref{eq:woPre}) (dashed line),
with the nuclear thickness and centrality quantilies determined from the
optical Glauber model (see \App{App} for details~\cite{Note2}). 
Both estimates are normalized to
reproduce the experimentally measured value of ${dN_{\rm ch}}/{d \eta}$ in
the $10-20 \%$ centrality class of Pb--Pb collisions at $\sqrt{s_\text{NN}}=2.76\,\text{TeV}$.
Different
trends in the centrality dependence of ${dN_{\rm ch}}/{d \eta}$ are clearly
visible, indicating the importance of the pre-equilibrium phase when comparing
observables of this type to experimental data.

Since ${dN_{\rm ch}}/{d \eta}$ is accurately measured in experiment for a wide
variety of colliding systems and energies, we can also compare the two
estimates directly to experimental data, which are reported as symbols in the
upper panel of \Fig{fig:Nvscent}. It is interesting to note that the average
$\left<{dN_{\rm ch}}/{d \eta}\right>$ as a function of centrality possesses a remarkable
degree of universality, such that---up to an overall normalization factor for
each collision system---data points for different colliding species
(Au, Cu, Pb, U, Xe) at RHIC and LHC energies all collapse onto a single curve to
high accuracy. Despite the simplicity of our theoretical estimate, we find that
the curve including pre-equilibrium effects provides a rather good description
of the experimental data, except for the more peripheral bins, where
fluctuations play an important role (see below). Due to the non-trivial
geometry dependence in \Eq{eq:wPre}, the calculation including pre-equilibrium
dynamics provides a much better description of the data than the estimate
in \Eq{eq:woPre}, which is based solely on the initial state.

Even though \Eq{eq:wPre} can be clearly justified from theoretical
calculations, our description is by no means unique.
Other phenomenological models \cite{Drescher:2006ca,Alver:2008aq,Gale:2012rq,Niemi:2015qia,Bernhard:2016tnd} successfully reproduce the centrality dependence seen in \Fig{fig:Nvscent}
by introducing various sources of  event-by-event fluctuations such as
 number and positions of participant nucleons,
their interaction strength, etc. 
However, it is important to emphasize in this
context that the pre-equilibrium phase also modifies the statistics of
fluctuations, such that for the long wavelength perturbations~\cite{Keegan:2016cpi,Kurkela:2018vqr}
\begin{eqnarray}
\label{eq:flucmod}
\frac{\delta s_{\rm hydro}}{s_{\rm hydro}} = \frac{2}{3}\frac{\delta e_{0}}{e_{0}},
\end{eqnarray}
which follows from the linearization of \Eq{eq:Entropy}. 
While
\Eq{eq:wPre} over-predicts particle production in peripheral collisions, it is
therefore not surprising that one can restore agreement with peripheral data by
including event-by-event fluctuations. This is demonstrated by the dot-dashed line in
\Fig{fig:Nvscent}, where the average of the nuclear thickness in \Eq{eq:wPre} has been
determined from a Glauber Monte Carlo model~\cite{Miller:2007ri}
(see \App{app} for details~\cite{Note2}).

\begin{figure}
  \centering
  \includegraphics[width=\columnwidth]{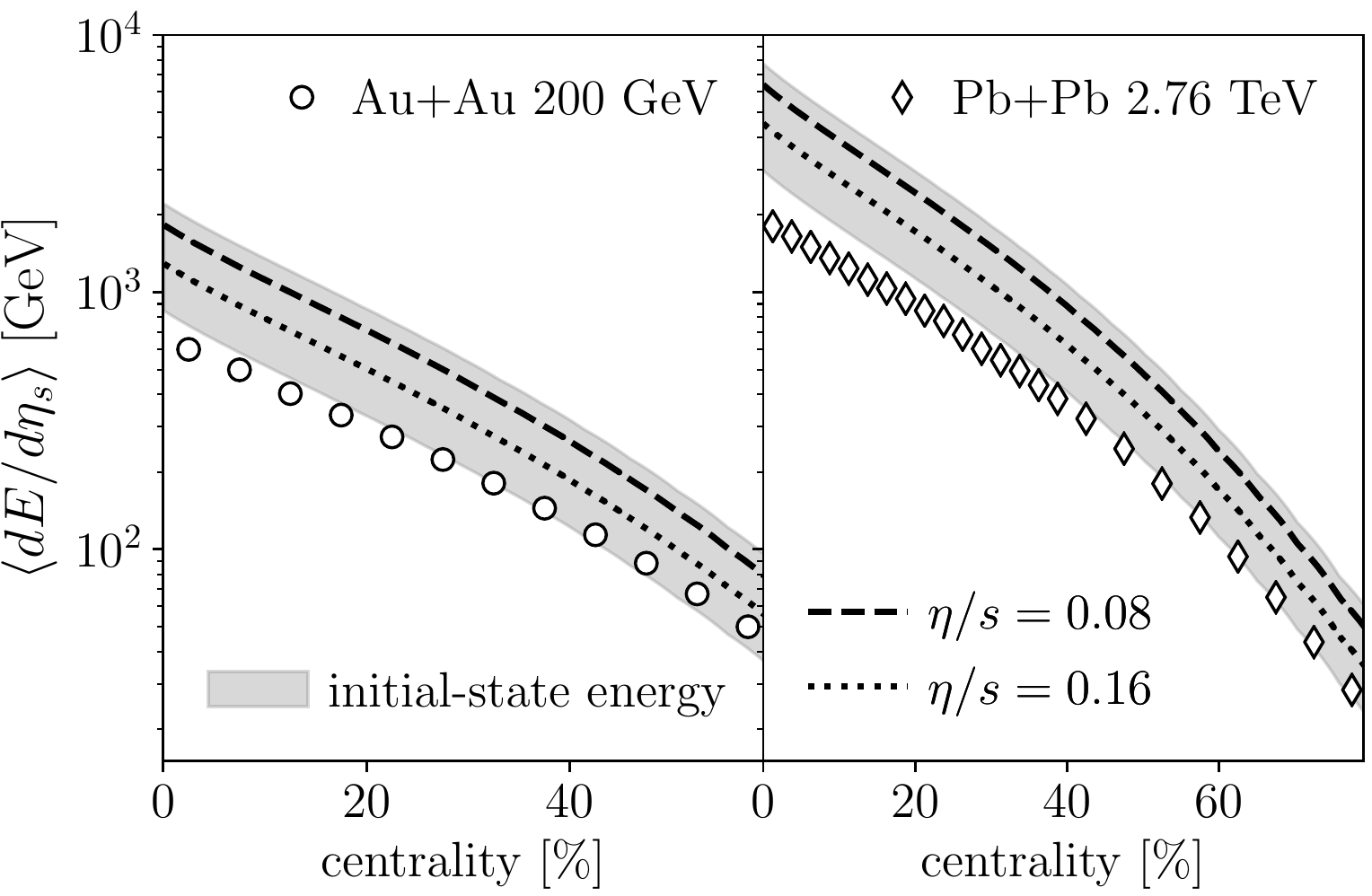}
  \caption{Estimate of the initial energy per unit space-time rapidity $dE_0 /d\etas$ (shaded area) determined from the measured particle multiplicity in $\sqrt{s_\text{NN}}=200\,\text{GeV}$ Au-Au (left) and
    $\sqrt{s_\text{NN}}=2.76\,\text{TeV}$ Pb-Pb collisions (right). Bands correspond to variations of $\eta /s=0.08\text{--}0.24$ and $\C=0.80\text{--}1.15$ while keeping all other parameters fixed as in \Eq{eq:InitialEnergyDensity}, and similarly the dashed lines correspond to specific values of $\eta/s=0.08,0.16$.  For comparison we show data points for the experimentally measured final state energy $dE_\text{final} /dy$~\cite{Adam:2016thv,Adare:2015bua},
    which is smaller due to the work performed against the longitudinal expansion.\label{fig:eng-pdf}}
\end{figure}

\prlsection{Estimating the initial-state energy density}%
So far we illustrated the utility of \Eq{eq:Entropy} for describing entropy
production  in the pre-equilibrium phase of high-energy
heavy-ion collisions. However, an equally important application concerns the
inverse problem, namely the estimation of the energy density $e_0$ of the
non-equilibrium state at very early time $\tau_0$ from experimental
measurements of hadrons in the final state. 

By inverting \Eqs{eq:Entropy} and \eqs{eq:Multiplicity} 
(and inserting typical values of
$dN_\text{ch}/d\eta\sim 1600$ and $A_\bot\approx \pi R_\text{Pb}^2\approx 138\,\text{fm}^2$)
we obtain the following estimate for the initial energy density 
for central Pb-Pb collisions at $\sqrt{s_{NN}}=2.76~{\rm TeV}$:
\begin{eqnarray}
&&e_{0} \approx 270~{\rm GeV / fm^3}~\left(\frac{\tau_0}{0.1 {\rm fm}/c}\right)^{-1} \left(\frac{C_{\infty}}{0.87}\right)^{-9 /8}
\!\left(\frac{\eta/s}{2/4\pi}\right)^{-1/2}  \nonumber
\\
&&
\left(\frac{A_{\bot}}{138 {\rm fm}^2}\right)^{-3/2} 
\left(\frac{dN_{\rm ch}/d\eta}{1600}\right)^{3/2} \left(\frac{\nu_{\rm eff}}{40}\right)^{-1/2} \left(\frac{S/N_{\rm ch}}{7.5}\right)^{3/2}\!\!\!\!,
\label{eq:InitialEnergyDensity}
\end{eqnarray}
at a time $\tau_0 = 0.1\,\text{fm}/c$, which should be at least of the order of
the formation time $1/Q_s \approx 0.1 \text{fm}/c$~\cite{Schlichting:2019abc}, but
small compared to the equilibration time $\tau_{\rm hydro} \ \approx 1\, {\rm
fm}/c$ \cite{Kurkela:2018wud,Kurkela:2018vqr} for the estimate in
\Eq{eq:InitialEnergyDensity} to be valid.
One finds that the initial
energy density quoted in \Eq{eq:InitialEnergyDensity}  is nearly three orders of magnitude higher than the energy
density at the QCD cross-over $e_{c} \approx 0.346(41) ~{\rm GeV}/\text{fm}^3$
(for 2+1 flavor QCD)~\cite{Ding:2015ona,Bazavov:2018mes}.

We emphasize that, unlike the usual Bjorken
estimate based on the measured final-state energy $e_0 = \frac{1}{\tau_0 A_\bot}
{dE_\text{final}}/{dy}$~\cite{Bjorken:1982qr}, our estimate in
\Eq{eq:InitialEnergyDensity} includes the work done during the
expansion of the QGP~\cite{Gyulassy:1983ub}. We demonstrate this effect in \Fig{fig:eng-pdf} where we
compare the experimentally measured $dE_\text{final} /d y$
in the final state
to the  initial-state energy per unit rapidity $dE_0/d\eta_s = \int
d\x_\perp (e\tau)_0$  reconstructed from the measured particle multiplicities as in \Eq{eq:InitialEnergyDensity}.
Note that to better account for the non-trivial
transverse geometry, we have estimated the transverse area $A_{\bot}$ from our fluctuating initial state model as described in \App{}~\cite{Note2}.

Based on this analysis, we find that, especially in central collisions at high
energies, the initial state $dE_0/d\etas$ can easily exceed the measured
${dE_\text{final}}/{dy}$ in the final state by a factor of two to three. Evidently
the exact amount of work done during the expansion
is subject to uncertainties in the
non-equilibrium and transport properties of the QGP, which we quantify by
uncertainty bands in \Fig{fig:eng-pdf}, corresponding to variations of
$\eta/s$ and $\C$ within the anticipated margins ($\eta/s=0.08-0.24$ and
$\C=0.80-1.15$). Vice versa, the size of the uncertainty bands in \Fig{fig:eng-pdf}
also demonstrates the fact that, if the initial state energy density can be
determined precisely, e.g.\ from theoretical calculations, then the
experimentally measured $dN_{\rm ch}/d\eta$ will impose strong constraints on the
non-equilibrium evolution of the QGP  characterized by $\C$ and $\eta/s$. Since, the $dE_0 /d\etas$ should always be larger than $dE_\text{final} /dy$ due to the work performed against the longitudinal expansion, one can further rule out values in the parameter space, where the initial $dE_0/d\etas$ falls below the experimental data points. Specifically, large values of $\eta/s$ and $\C$ can already be ruled out, because the estimated  $dE_0/d\etas$ in peripheral collisions turns out to be unphysically small, i.e. below the experimental points of $dE_\text{final}/dy$, which provide a lower bound on
the initial state energy.

\prlsection{Discussion}%
Entropy production in high-energy heavy-ion collisions occurs predominantly during the earliest stages, when the system is significantly out-of-equilibrium; therefore measurements of the charged particle multiplicities---reflecting the total amount of entropy produced in the collision---provide a highly sensitive probe of the pre-equilibrium dynamics. Based on the concept of hydrodynamic attractors, which give a macroscopic description of the early time dynamics of the QGP, we established for the first time a direct relation between the initial-state energy and the final-state entropy. This relation, \Eq{eq:Entropy}, is remarkably insensitive to the microscopic details of the approach to equilibrium (see \Fig{fig:attr}).

By combining the information from $dN_\text{ch}/d\eta$ on entropy production and $dE_\text{final} /dy$ on the work performed against the longitudinal expansion, we demonstrate
that a precise calculation of the initial state energy can impose stringent constraints on the shear-viscosity to entropy density ratio $\eta/s$.
Based on our extraction of $dE_0/d\etas$, which assumes a scenario of (nearly-)complete equilibration, we obtain an upper limit for $\eta/s \lesssim 0.4$ for the most favorable choice of all other parameters.
Conversely, for $\eta/s \gtrsim 0.4$ we can not expect the QGP to equilibrate in peripheral nucleus-nucleus collisions (see also \cite{Bhalerao:2005mm, Kurkela:2019kip}) and our estimates need to be revised. Non-trivial modifications due to incomplete equilibration will arise in this context, which should be investigated further, for instance  by means of the \kompost{}  pre-equilibrium package~\cite{Kurkela:2018vqr,Kurkela:2018wud,kompost_github}. 
We expect such effects to become particularly important in collisions of smaller nuclei, e.g., p--A or  O--O, which may therefore provide even deeper insights into the fascinating out-of-equilibrium dynamics of the QGP.

\prlsection{Acknowledgments}We thank 
A.~Andronic,
J.~Berges,
N.~Borghini,
M.~Heller, 
C.~Klein-B\"{o}sing, 
A.~Kurkela,
T.~Lappi,
M.~Martinez,
J.Y.~Ollitrault,
K.~Reygers,
C.~Schmidt,
D.~Teaney,
and
R.~Venugopalan
for stimulating
discussions and P.~Romatschke for also providing 
the AdS/CFT attractor curve.
This work was supported in part by the German Research Foundation
(DFG) through the Collaborative Research Centres ``CRC-TR 211:
Strong-interaction matter under extreme conditions'' (S.S.) and ``SFB 1225: Isolated quantum systems and universality in extreme conditions (ISOQUANT)" (A.M.). We thank the Institute for Nuclear Theory at the University
of Washington for its kind hospitality and stimulating research environment.

\bibliography{main}

\appendix

\aprlsection{Supplemental material\label{sec:supplemental}}
Below we explain the details of the implementation of the initial state model described in the main text. When referring to the optical Glauber calculations, which yield the dashed and solid lines in \Fig{fig:Nvscent}, we determine the nuclear thickness $T(\mathbf{x}_\bot)
$ of each nucleus according to the longitudinal integral of the (average) nuclear matter
density distribution $T(\mathbf{x}_\bot) =\int_z dz~\rho(\sqrt{\mathbf{x}_\bot^2 - z^2 })$. We follow previous works and parametrize 
 $\rho(r)$ by a two-parameter Fermi distribution
\begin{equation}
\rho(r) = \rho_0 \biggl [1~+~\exp \biggl (
\frac{1}{a}\bigl[r - R \bigr ] \biggr )
\biggr]^{-1}\;,
\end{equation}
with $a=0.55$ fm and $R=6.62$ fm for $^{208}$Pb nuclei,
and $a=0.53$ fm, $R=6.40$ fm for $^{197}$Au nuclei~\cite{DeJager:1987qc}, while $\rho_0$ is such that $\rho(r)$ is normalized to the total number of nucleons.
The thickness functions of two nuclei (A and B) colliding at a finite impact parameter, $\mathbf{b}$, are given by $T^{\text{A}/\text{B}}(\mathbf{x}_\bot) = T(\mathbf{x}_\bot
\pm  \mathbf{b}/2)$. The centrality of a given collision is obtained from the geometric relation $centrality=\pi
|\mathbf{b}|^2/\sigma_\text{tot}$~\cite{Teaney:2009qa}, where $\sigma_\text{tot}$ is the total inelastic nucleus-nucleus
cross section, and we use $\sigma_\text{tot}=767$ fm$^2$ for Pb-Pb
collisions, and $\sigma_\text{tot}=685$ fm$^2$ in Au-Au collisions.

When referring to fluctuating initial conditions, which yield the dot-dashed line in \Fig{fig:Nvscent}, we introduce nucleonic degrees of freedom following the Monte Carlo version of the Glauber model~\cite{Miller:2007ri}. In both colliding nuclei we sample the positions of the nucleons according to $\rho(r)$, which we then shift by $\pm\mathbf{b}/2$ to take into account the finite impact parameter. Subsequently, we determine the \textit{participants} in a black-disk approximation as those nucleons of nucleus A that are located within a distance $\sqrt{ \sigma_{\rm nn} /\pi}$ from at least one nucleon in nucleus B, and vice versa. Here  $\sigma_{\rm nn}$ is the nucleon-nucleon cross-section, which is equal to $4.2$ fm$^2$ for $\sqrt{s_{\rm NN}}=200$ GeV and to $6.4$ fm$^2$ for  $\sqrt{s_{\rm NN}}=2.76$ TeV. The thickness $T(\mathbf{x}_\bot)$ of each nucleus is then determined on an event-by-event basis by summing the density profiles of all its participating nucleons, where the density of each participant nucleon is taken as a Gaussian profile of width $0.5$~fm.

Explicitly restoring the constants in Eqs.~(10) and (8), the saturation scale $Q_{s}(\mathbf{x})$ of the nucleus is determined as
\begin{eqnarray}
Q_{s}^{2}(\mathbf{x_\perp}) = Q_{sp}^{2}(\sqrt{s}) A_{p}~T(\mathbf{x_\perp})\;,
\end{eqnarray}
where $Q_{sp}(\sqrt{s})$ and $A_{p}=\pi R_{p}^2$ denote the saturation scale and (transverse) size of a proton, and the initial state energy density per unit rapidity is then given by
\begin{equation}
\label{eq:eee}
    (e\tau)_0(\mathbf{x}_\bot) = \mathcal N_{e}~\left(Q_{sp}^{2}(\sqrt{s}) A_{p}\right)^{3/2}~\,T^{<}(\mathbf{x}_\bot) \sqrt{T^{>}(\mathbf{x}_\bot)},
\end{equation}
where $\mathcal N_e$ is a dimensionless proportionality coefficient (see e.g.~\cite{Dumitru:2001ux}). By use of Eqs.~(6) and (7), the multiplicity in an event is then given by
\begin{align}
\frac{dN_\text{ch}}{d\eta} &= \frac{4}{3J} C_{\infty}^{3/4} \left(4\pi \frac{\eta}{s} \right)^{1/3} \left( \frac{\pi^2}{30} \nu_{\rm eff} \right)^{1/3}~\frac{N_{\rm ch}}{S} \nonumber \\
&\mathcal N_e^{2/3}~Q_{sp}^{2}(\sqrt{s}) A_{p}~\int d \mathbf{x}_{\bot}~  \left(T^{<}(\mathbf{x}_\bot) \sqrt{T^{>}(\mathbf{x}_\bot)}\right)^{2/3}, \label{eq:dNdeta2}
\end{align}
while the initial-state energy per unit rapidity is simply given by the transverse integral of $(e\tau)_0$ in Eq.~(\ref{eq:eee}) as
\begin{align}
\label{eq:InitialEnergySophisticated}
\frac{dE_0}{d\etas} =  \mathcal N_{e}~\left(Q_{sp}^{2}(\sqrt{s}) A_{p}\right)^{3/2}~\int d \mathbf{x}_{\bot}~  T^{<}(\mathbf{x}_\bot) \sqrt{T^{>}(\mathbf{x}_\bot)}.
\end{align}
Statistical averages of $\frac{dN_\text{ch}}{d\eta}$ and $\frac{dE_0}{d\etas}$ at a given centrality percentile are then obtained by sampling many events at the corresponding impact parameter. By matching the multiplicity in Eq.~(\ref{eq:dNdeta2}) to experimental data, we find that the dimensionless combination of pre-factors 
\begin{equation}
\frac{4}{3J} C_{\infty}^{3/4} \left(4\pi \frac{\eta}{s} \right)^{1/3} \left( \frac{\pi^2}{30} \nu_{\rm eff} \right)^{1/3}~\frac{N_{\rm ch}}{S}\mathcal N_e^{2/3}~Q_{sp}^{2}(\sqrt{s}) A_{p}
\end{equation}
in Eq.~(\ref{eq:dNdeta2}) should be equal to $10.7$ for 2.76 TeV Pb-Pb collisions $4.8$ for 200 GeV Au-Au collisions, which is compatible with the expected increase of the saturation scale $Q_{s}^{2}(\sqrt{s}) \propto \sqrt{s}^{\lambda}$ with $\lambda \approx 0.3$~\cite{Albacete:2014fwa}.

Vice versa, to determine the initial state energy per unit rapidity shown in \Fig{fig:eng-pdf}, we can use \Eq{eq:dNdeta2} for the multiplicity to eliminate the pre-factor $N_{e}~\left(Q_{sp}^{2}(\sqrt{s}) A_{p}\right)^{3/2}$ from \Eq{eq:InitialEnergySophisticated} and express $\frac{dE_0}{d\etas}$ in the form of Eq.~(\ref{eq:InitialEnergyDensity}) as
\begin{align}
\label{eq:dEdetafinal}
\frac{dE_{0}}{d\etas}&= \left(\frac{4}{3J} \right)^{-3/2}  C_{\infty}^{-9/8} \left(4\pi \frac{\eta}{s} \right)^{-1/2} \left( \frac{\pi^2}{30} \nu_{\rm eff} \right)^{-1/2} \nonumber \\
&~A_{\bot}^{-1/2} \left( \frac{S}{N_{\rm ch}} \right)^{3/2} \left(\frac{dN_\text{ch}}{d\eta}\right)^{3/2},
\end{align}
where the effective area, $A_\bot$, is given by
\begin{align}
\label{eq:fancyArea}
\sqrt{A_{\bot}}=\frac{\left(\int d \mathbf{x}_{\bot}~ \left(T^{<}(\mathbf{x}_\bot) \sqrt{T^{>}(\mathbf{x}_\bot)}\right)^{2/3}\right)^{3/2}}{\int d \mathbf{x}_{\bot}~  T^{<}(\mathbf{x}_\bot) \sqrt{T^{>}(\mathbf{x}_\bot)}}.
\end{align}
Evaluating Eq.~(\ref{eq:fancyArea}) within our Monte-Carlo calculations yields $A_\bot \approx 120$ fm$^2$ in central Pb-Pb collisions, which is in fair agreement with the value anticipated in Eq.~(\ref{eq:InitialEnergyDensity}). Note that, while we could have used the experimental data for $dN_{\rm ch}/d\eta$ to produce \Fig{fig:eng-pdf}, we conveniently make use of the model estimate in Eq.~(\ref{eq:dNdeta2}) to interpolate between experimental data points; as shown in \Fig{fig:Nvscent} the model is within $5\%$ of all experimental data.

\end{document}